\documentclass[final,3p,times,twocolumn]{elsarticle}

\usepackage{graphics}

\usepackage{amssymb}

\journal{Special Issue of Fusion Engineering and Design}

\newcommand{\ig}[1]{\includegraphics[width=8cm]{#1}}

\begin{document}

\begin{frontmatter}

\title{New magnetic real time shape control for MAST}

\author[label1]{L. Pangione}
\author[label1]{G. McArdle}
\author[label1]{J. Storrs}
\address[label1]{EURATOM/CCFE Fusion Association - Culham Science Centre, Abingdon, Oxon, OX14 3DB, UK}

\begin{abstract}
The MAST (Mega Ampere Spherical Tokamak) real time plasma position controller is based on an optical linear camera placed on the mid plane of the vessel. This solution has the advantage of being a direct observation of the $D_\alpha$ emissions coming from the interaction between the boundary of the plasma and neutral gas, but, on the other hand, it restricts the control to the outer radius of the plasma only. 

A complete chain of tools has been set up to implement, test and simulate a new real time magnetic plasma shape controller based on the rtEFIT code. The complete working path consists of three elements: a linear static relationship between control parameters and current demands, a linear state space model needed to represent the plasma dynamic response in closed loop simulations, and the possibility to run simulations inside the Plasma Control System (PCS). The linear relationship has been calculated using the FIESTA code, which is developed using Matlab at CCFE. The linear state space model was generated using the CREATE-L code developed by the CREATE Consortium. It has already been successfully used to model JET, FTU and TCV tokamaks. Using this working path many simulations have been carried out allowing fine tuning of the control gains before the real experiment. The simulation testing includes the plasma shape control law as implemented in PCS itself, so intensive debugging has been possible prior to operation. Successful control using rtEFIT was established in the second dedicated experiment during the MAST 2011-12 campaign. This work is a stepping stone towards divertor control which is ultimately intended for application to the super-X divertor in the MAST Upgrade experiment.

\end{abstract}


\end{frontmatter}

\section{Introduction}
\label{ref:intro}

The MAST real time plasma shape control system is currently based on an optical linear camera placed on the mid plane of the vessel, capturing $D_\alpha$ emissions in a single row of $2048$ pixels \cite{homer}. The camera controller computes the plasma outer radius and sends it through an analogue link to the main plasma shape control system. This value is processed to produce current demands. The relationship between plasma position and demands is a primitive linear mapping. This solution has the advantage of being fast and insensitive to magnetic noise, but shows its limits when considering MAST Upgrade and in particular its super X divertor requirements \cite{mastuSxd}. Though initial optical control may be used initially in MAST Upgrade, it will not be applicable for the control of the MAST Upgrade plasma shape and its super X divertor region. 

A new real time magnetic shape controller for the MAST experiment has been successfully implemented and tested during the MAST 2011-12 campaign. 

In this paper we describe the work done to produce a complete chain of tools that allowed us to implement a new magnetic shape controller for MAST experiment. In section \ref{ref:chPcs} we report a brief introduction of the MAST plasma control system and the tools needed to prepare the magnetic shape controller. In section \ref{ref:chSimEnv} we present modifications to the existing simulation environment to produce closed loop simulations and their results. In section \ref{ref:chExpRes} we report the results obtained in dedicated experimental sessions during the MAST 2011-12 campaign.

\section{Plasma Control System}
\label{ref:chPcs}
\begin{figure}
\ig{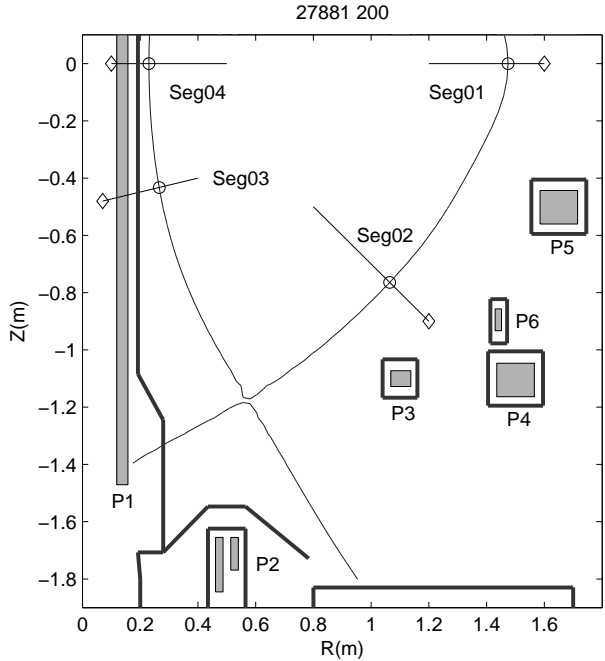}
\caption{Position of the control segments and control points with respect of a typical MAST plasma.}
\label{fig:mastSection}
\end{figure}

MAST uses the Plasma Control System (PCS) \cite{pcsMAST} developed by General Atomics, which is currently used in many tokamak devices around the world \cite{pcsWorldWide}. It offers a complete environment needed to run a machine: the real time infrastructure, the data storage system, the graphical user interface (GUI) used by the session leaders to run the experiments, etc. It also provides a real time equilibrium reconstruction code (rtEFIT) \cite{Ferron01} that can be customized to run on the specific tokamak device. The rtEFIT based real time shape controller uses the isoflux technique, which constrains the values of the flux at predefined points on control segments to be equal to the value of the flux at the reference point. By choosing the reference point to be a location known to be on the plasma boundary (e.g. the limiter touchpoint or the X-point of a divertor plasma), this method will control the boundary flux contour to pass through all the pre-defined control points. In figure \ref{fig:mastSection} is shown the position of the control segments and control points in relation with a typical MAST plasma boundary. The figure also shows the position of the solenoid ($P1$) and of the poloidal coils ($P2$, $P3$, $P4$, $P5$, $P6$). It is important to note that in this paper we will not consider $P3$ coil because it is used only during the creation of the plasma and $P6$ coil because it is used only for the vertical stabilization. 

\begin{figure}
\ig{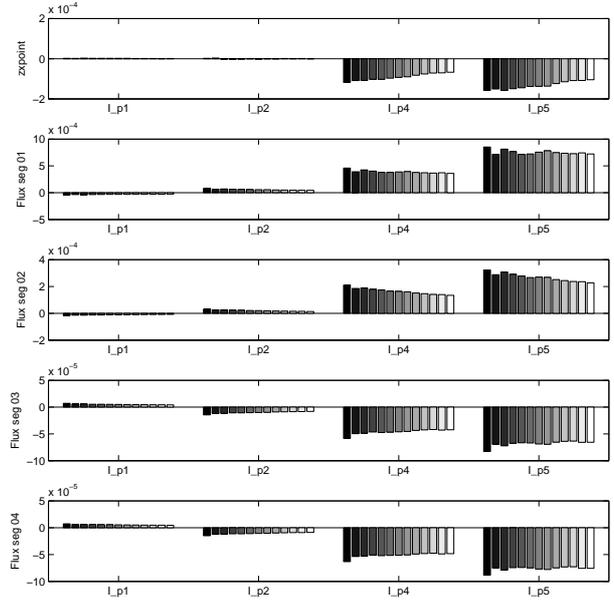}
\caption{Evolution of the sensitivity matrices during the plasma current flat-top for a preparation shot.}
\label{fig:smats}
\end{figure}

From the GUI, there are many options for configuring the required position and shape evolution during the shot, but all of these are converted to equivalent control point references. During the discharge rtEFIT reconstructs the equilibrium and produces flux and geometrical errors at control points as outputs. The machine specific control law corrects these errors by driving them to zero. The control law mainly consists in a relation that maps the flux errors into coil current commands and a proportional controller. The simplest relation is a linear and static map $\Delta p = S \cdot \Delta I$, where $S$ is defined as sensitivity matrix, $\Delta I$ is the variation of the coil currents from the starting equilibrium and $\Delta p$ is the variation of the parameters vector. Considering the flux value on the reference positions and the geometrical $Z$ position of the X point as parameters, the current variations needed to correct a $\Delta p$ error can be simply obtained by inverting the $S$ matrix, so $\Delta I = S^{-1} \cdot \Delta p$. Clearly this simple relation is valid only for small parameter errors. The pseudo (or the pseudo-inverse) $S^{-1}$ depends on the control set up, i.e. control segments and activated control coils, and is pre-calculated before the shot. Figure \ref{fig:smats} shows an offline calculation of the sensitivity matrices during the plasma current flat-top for a preparation shot. The values of the matrices do not change significantly, so it is possible to consider only one sensitivity matrix for the entire duration of the plasma current flat-top. The plasma current ramp up and termination phases will use different sensitivity matrices and a similar algorithm in which the considered reference point is the touching limiter point. These experiments are planned for MAST 2013 experimental campaign.

The sensitivity matrix is calculated using the FIESTA code. FIESTA is a Grad Shafranov free boundary solver developed in Matlab by Dr. G. Cunningham at Culham Centre for Fusion Energy (CCFE). Starting from an equilibrium obtained by an EFIT reconstruction of a discharge at a given time, all the coil currents are independently perturbed and the resulting variation in the parameters is computed. The sensitivity value is the slope of the linear approximation relation between perturbed parameters and perturbed currents.

\section{Simulation environment}
\label{ref:chSimEnv}
The rtEFIT code is complex and has a large number of settings and parameters to be defined in the compilation phase and during the shot preparation phase. A significant effort was invested to ensure that the calculations remain valid throughout the discharge. Also, many parameters, such as the proportional control gains, needed to be tuned before the experiments.

PCS already offers a simulation environment that replaces the hardware I/O commands with software messages from/to a simulation server (simserver). The existing MAST implementation only supported replay of MAST data for open loop testing but ignored controller actions.

This can be improved by the use of a linear time invariant (LTI) model of the plasma which reacts to the commands. A plasma linear model for CCFE has been provided by the CREATE consortium through the XSCtools, which produce an LTI state space model representing the plasma and the machine at a given time for a given shot. These tools have successfully been used in many other tokamak devices like JET \cite{createJet}, FTU, TCV  and are key tools for the design of ITER. A CREATE LTI model for MAST returns simulated data for all the magnetic sensors available in the experiment as well as a geometrical estimate of the plasma boundary. A comprehensive evaluation of the CREATE LTI model for MAST experiment, without using the PCS simulation environment, is reported in \cite{Artaserse01}.

Given a model of the plasma, it was possible to modify the PCS simulation environment to replace, from a predefined time $t_{switch}$, archived data with data coming from a running Matlab/Simulink process. In this way the part of the simulation before the $t_{switch}$ is open loop while after the $t_{switch}$ is closed loop. Synchronous two-way communication between the PCS simserver and the Matlab process is guaranteed by POSIX queues. It is important to remark that the main purpose of this simulation environment is not to predict the exact behaviour of the plasma and the machine, but only to give the possibility to tune the parameters. The main constraint is represented by the nonlinearity of the plant, like for example saturations in voltages and currents and time delays, which cannot be represented by a linear model. 

\begin{figure}
\ig{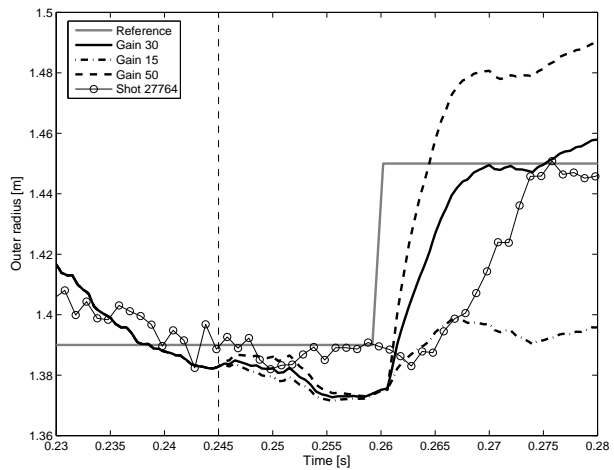}
\caption{Outer radius simulation results compared with the experimental data. An optimal proportional gain is reported in black solid line, a lower gain in dash-dotted black solid line and an higher gain in dashed black line. Solid dark grey represents outer radius reference, while solid-circle light grey is the actual experiment outer radius. The vertical dashed line represents the beginning of the closed loop simuation $t_{switch}$.}
\label{fig:simOutputs}
\end{figure}

Using this simulation environment, a lot of development and testing has been done. An example of simulation output is reported in figure \ref{fig:simOutputs}, which shows the behaviour of the outer radius control action for different values of proportional gain. As can be seen a good choice of the gain to follow the step change is 30. A lower gain does not allow the system to react to the perturbation, while a higher gain brings the system close to instability. It is important to observe that the model and the experimental data needed for the open loop part of the simulation were from a MAST experiment in which rtEFIT was deactivated. The PCS configuration set in the closed loop part of the simulation have been then used in a real MAST experiment. As can be seen there is a good agreement between the simulated, solid black line, and the actual behaviours, solid-circle light grey line. Considering the good agreement previously obtained in the CREATE LTI evaluation without the PCS simulation environment, the discrepancy in response time is probably due to the model of the power supplies; an improved model of them is still in development. 

\section{Experimental results}
\label{ref:chExpRes}

Dedicated experiments have been carried out during MAST 2011-12 campaign. In order to optimize allocation of the machine time, it was decided to run short sessions of two or three plasma shots at intervals of a week. This approach allowed us to check results carefully and design the next experiments. It is important to remark that, in any case, successful control of the plasma shape was established in the second shot.

\begin{figure}
\ig{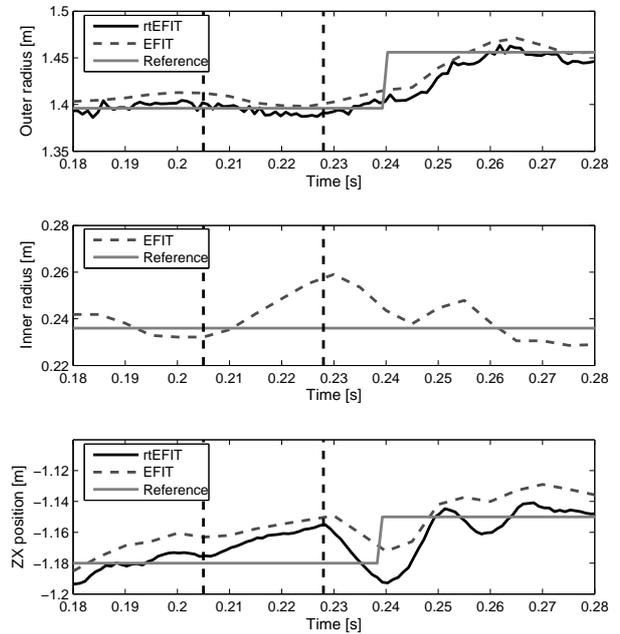}
\caption{Main geometrical measurements for MAST experiment number 28026. Outer radius, inner radius and $Z$ position of the X point are shown. rtEFIT data are reported in black line, EFIT in dark grey dashed line and references in light grey. Vertical dashed lines represent the beginning and the end of P2 inactivity.}
\label{fig:exp_ctrlErr}
\end{figure}

\begin{figure}
\ig{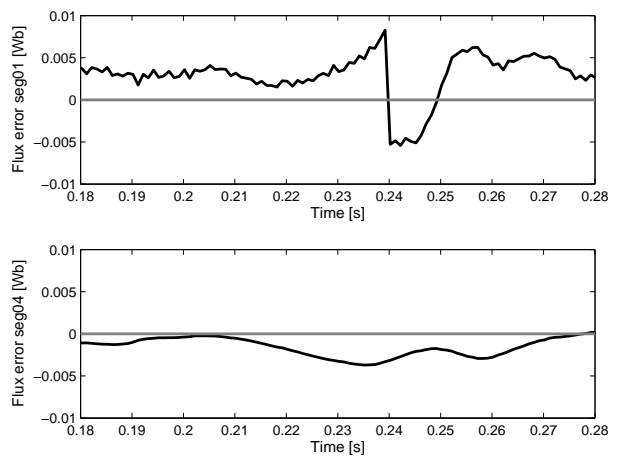}
\caption{Flux errors for MAST experiment 28026.}
\label{fig:exp_segErr}
\end{figure}

Main results of the experiments are reported in figures \ref{fig:exp_ctrlErr} and \ref{fig:exp_segErr}. 
In particular in figure \ref{fig:exp_ctrlErr}, three main geometrical plasma parameters, outer radius, inner radius and $Z$ position of the X point, are reported. For each of them an independent measurement obtained from EFIT data is also shown. As can be seen, the three measurements correctly follow the references except between $0.205 s$ and $0.228 s$, when P2 power supply was disabled to allow operation of the polarity reversing switch. Once this reversal finished, the P2 coil is back in control and correctly drives the $Z$ position of the X point. In figure \ref{fig:exp_segErr} the flux errors of the two active control segments are reported. As can be seen, the reference variation of the major radius of the plasma suddenly changes the flux errors of segments $1$ and $4$.

It is important to note that the controller is not aware P2 power supply reversing delay. A more sophisticated implementation could change the active sensitivity matrix according to the state of the power supplies. This will developed if needed in the MAST Upgrade plasma shape control system. 

\section{Conclusions}
\label{ref:chConc}

A new real time magnetic shape controller for the MAST experiment has been successfully implemented and tested during the MAST 2011-12 campaign. Successful control using rtEFIT was established in the second dedicated experiment during the MAST 2011-12 campaign. This good achievement has been possible also thanks to the multiple tools available at MAST. This work is a stepping stone towards divertor control which is ultimately intended for application to the super-X divertor in the MAST Upgrade experiment.

\section{Acknowledgements}
\label{ref:chAck}

This work was funded by the RCUK Energy Programme under grant EP/I501045 and the European Communities under the contract of Association between EURATOM and CCFE. The views and opinions expressed herein do not necessarily reflect those of the European Commission.

Authors want to acknowledge CREATE Consortium team, in particular Prof Albanese, Dr G Artaserse and Dr F Maviglia, and the General Atomics team for their support.

\end{document}